\documentclass{ifacconf}

\usepackage{graphicx}      
\usepackage{xcolor}
\usepackage{natbib}       
\usepackage{booktabs}
\usepackage{epsfig} 
\usepackage{mathptmx} 
\usepackage{times}
\DeclareMathAlphabet{\mathcal}{OMS}{cmsy}{m}{n} 
\usepackage{amsmath}
\usepackage{amssymb} 
\usepackage{subcaption}
\newtheorem{theorem}{Theorem}
\newtheorem{corollary}{Corollary}[section]
\newtheorem{lemma}{Lemma}[section]

\newtheorem{definition}{Definition}[section]

\begin{document}
\begin{frontmatter}

\title{An Input-Output Data-Driven Dissipativity Approach for Compositional Stability Certification of Interconnected LTI MIMO Systems}

\author[BTU]{Alejandra Sandoval-Carranza}
\author[BTU]{Juan E. Machado}
\author[BTU,IEG]{Johannes Schiffer}

\address[BTU]{Brandenburg University of Technology Cottbus-Senftenberg,
03046 Cottbus, Germany (e-mail: \{sandoval, machadom, schiffer\}@b-tu.de).}

\address[IEG]{Fraunhofer IEG, Fraunhofer Research Institution for Energy Infrastructures 
and Geotechnologies IEG, 03046 Cottbus, Germany\\ 
(e-mail: johannes.schiffer@ieg.fraunhofer.de).}

\begin{abstract}              
We propose an input-output data-driven framework for certifying the stability of interconnected multiple-input-multiple-output linear time-invariant discrete-time systems via QSR-dissipativity. 
That is, by using measured input-output trajectories of each subsystem, we verify dissipative properties and extract local passivity indices without requiring an explicit model identification.
These passivity indices are then used to derive conditions under which the equilibrium of the interconnected system is stable.
In particular, the framework identifies how the lack of passivity in some subsystems can be compensated by surpluses in others.
The proposed approach enables a compositional stability analysis by combining subsystem-level conditions into a criterion valid for the overall interconnected system. 
We illustrate via a numerical case study, how to compute channel-wise passivity indices and infer stability guarantees directly from data with the proposed method.

\end{abstract}


\end{frontmatter}

\section{Introduction}

\subsection{Motivation}

In many modern infrastructures—such as power systems, transportation and communication networks—there exist complex interdependencies that can compromise overall stability and reliable operation: in power systems, the rapid integration of renewable and distributed resources intensifies dynamic couplings and reduces inertia, fundamentally challenging traditional notions of stability~\citep{laib2023decentralized,cucuzzella2024microgrids,machado2023online,schiffer2016Asurvey,schiffer2019OnlineEstimation}; in formation control of mobile robots, changes in the communication or sensing graph can cause formation errors or instability~\citep{fax2004information}.
Similar challenges appear across multi-agent and networked control systems more broadly, where heterogeneity, subsystem coupling, and limited measurements complicate stability assessment~\citep{weiland2022interconnection,hill2022dissipativity}. In such settings, compositional criteria are more beneficial and easier to apply than centralized ones~\citep{martinelli2024interconnection,murat2022compositional}. Moreover, there is an emerging interest in moving from traditional model-based analysis to data-driven concepts due to the growing ease of obtaining and processing data as well as the increase in system complexity~\citep{depersis2019formulas,van2023behavioral}.

\subsection{Literature Review}

Existing stability assessment frameworks for interconnected systems are typically model-based~\citep{machado2020passivity,martinelli2024interconnection} and assume diffusive (e.g., Laplacian-type) interconnection structures~\citep{malan2025passivation,martinelli2024interconnection}. Within the available literature, dissipativity and passivity theories provide a rigorous foundation for compositional analysis by quantifying energy exchange between a subsystem and the environment in which it is embedded~\citep{willems1972dissipative,lozano2013dissipative}. In particular, passivity indices, which can quantify a system's deviation from passivity~\citep{khalil2002nonlinear}, have enabled the provision of decentralized stability guarantees and guided controller design in networked settings~\citep{he2024passivity,malan2023passivitybased}. However, both dissipativity and passivity assessment approaches conventionally rely on explicit system models, limiting their applicability in large-scale or uncertain settings, see~\cite{machado2020passivity,strehlemachado2024TCST,martinelli2024interconnection}.

To address this issue, more recently data-based approaches have been proposed for verifying dissipativity properties directly from measured system trajectories, see~\cite{waarde2022slemma,koch2022robust,van2023informativity,markovsky2021behavioral}. Most of the available solutions that can be tailored for the analysis and control of MIMO systems are input-state approaches, i.e., they rely on full-state measurements, see~\cite{depersis2019formulas,van2023behavioral,nakano2025dissipativity}. However, in many applications obtaining full state measurements is infeasible, especially in large-scale systems. Therefore, data-driven input–output methods that allow for dissipativity assessment based solely on input and output trajectories are more appealing. Yet, existing methods are limited in the sense that formulations assume zero initial conditions~\citep{maupong2017lyapunov,rosa2021one}, focus on SISO settings~\citep{koch2022robust,depersis2019formulas}, and their application to MIMO interconnected systems remains limited~\citep{alsalti2023notes,verhoek2024lpv}.

\subsection{Contributions}

In light of the above discussion, the main contributions of this work are summarized as follows:
\begin{itemize}
    \item By using an input-output data-driven approach, we report an LMI-based characterization of QSR-dissipativity for discrete-time LTI MIMO systems, without resorting to stringent assumptions on the systems initial conditions. 
     \item  By introducing the notion of \emph{channel-wise} passivity indices, we exploit the proposed data-based QSR-dissipativity assessment approach for deriving a flexible compositional criterion for investigating the stability of interconnected LTI MIMO systems based solely on local input-output data, i.e., without relying on centrally available system data or measurements.
    \item Via a case study based on a realistically parameterized DC microgrid model, we illustrate the applicability and interpretability of the proposed data-based compositional stability criterion, allowing us to visualize the effect of faults in energy generation units on the channel-wise passivity indices and the overall system stability.
\end{itemize}

\subsection{Paper Organization}
The article is organized as follows. 
In Section~II, we provide the necessary preliminaries and notation. 
In Section~III, we state the problem and describe the considered system interconnection structure. 
In Section~IV, we develop the compositional stability analysis based on QSR-dissipativity and channel-wise passivity indices. 
In Section~V, we illustrate the results on a numerical DC microgrid example. 
In Section~VI, we conclude the paper.

\subsection{Notation}
We denote the set of integer numbers by $\mathbb Z$ and the set of real numbers by $\mathbb R.$
For a signal $z : \mathbb Z \to \mathbb{R}^n$ and two integers $i$ and $k$, with $i\leq k$, we employ the notation $z_{[i,k]} := \{z(i), z(i + 1), \ldots , z(k)\}.$
For a signal $z$ and a positive integer $T$, we follow \cite{depersis2019formulas} and define the Hankel matrix associated to $z$ as: 
\begin{equation}
Z_{\{i,t,T\}}=
    \begin{bmatrix}
        z(i) & \cdots & z(i+T-1)\\
        \vdots & \ddots & \vdots\\
        z(i+t-1) & \cdots & z(i+t+T-2)
    \end{bmatrix} \in \mathbb{R}^{tn\times T},
\end{equation}
where the subscript $i$ denotes the time at which the first sample of the signal is taken, $tn$ the number of rows and $T$ the number of columns. When $t=1$, we denote~\citep{depersis2019formulas}: 
\begin{equation}
    Z_{\{i,T\}}:=\begin{bmatrix}
        z(i) & z(i+1) & \cdots & z(T+i-1)
    \end{bmatrix} \in \mathbb{R}^{n \times T}.
    \label{eq:Z}
\end{equation}
For a matrix $A\in\mathbb{R}^{n\times m},$ $A^\dagger$ denotes its Moore-Penrose inverse. 

\section{Preliminaries}\label{sec:preliminaries}

In this section, we introduce instrumental concepts and  results  about QSR-dissipative systems and data-based characterization of discrete-time linear time-invariant (LTI) multiple-input-multiple-output (MIMO) systems.

\subsection{QSR-Dissipativity}\label{sec:data_based}

We consider the following discrete-time LTI MIMO system:
\begin{equation}
    \begin{split}
        x(k+1)&=Ax(k)+Bu(k),\\
        y(k) &= Cx(k),
    \end{split}
    \label{eq:originalsystem}
\end{equation}
where $x \in \mathbb{R}^{n}$, $u \in \mathbb{R}^{m}$, and $y \in \mathbb{R}^{p}$ are the state, input and output vector, respectively. A standing technical assumption throughout this section is that the system~\eqref{eq:originalsystem} is controllable and observable.

We employ the following standard notions of dissipativity and QSR-dissipativity \citep{Byrnes_losslessnes1994,martinelli2024interconnection}.
\begin{definition}[Dissipativity]
The system \eqref{eq:originalsystem} is said to be \emph{dissipative} with respect to the supply rate 
$w: \mathbb{R}^m \times \mathbb{R}^{p}\rightarrow \mathbb{R}$ if there exists a nonnegative function 
$V: \mathbb{R}^{n}\rightarrow \mathbb{R}$, called a storage function, such that for all 
$u \in \mathbb{R}^{m}$, all $k\in \mathbb{N}$ and any initial condition $x(0)\in\mathbb{R}^n$,
\begin{equation}
     V(x(k+1))-V(x(k))\leq w(u(k),y(k)).
     \label{eq:dissipative_inequality}
\end{equation}
\end{definition}

\begin{definition}[QSR-dissipativity]
The system \eqref{eq:originalsystem} is said to be \emph{QSR-dissipative} if it is dissipative with respect to the supply rate
\begin{equation}
    w(u(k), y(k)) = 
    \begin{bmatrix} y(k) \\ u(k) \end{bmatrix}^{\top}
    \begin{bmatrix} Q & S \\ S^{\top} & R \end{bmatrix}
    \begin{bmatrix} y(k) \\ u(k) \end{bmatrix},
    \label{eq:supply-qsr}
\end{equation}
where $Q=Q^{\top}\in\mathbb R^{p\times p}$, $S\in\mathbb R^{p\times m}$ and $R=R^{\top}\in\mathbb R^{m\times m}$.
\end{definition}

An LMI-based necessary and sufficient condition for QSR-dissipativity is as follows~\citep{moylanandhillstability,koch2022robust}.
\begin{theorem}\label{theorem:qsr_dissip_model_based}
The system \eqref{eq:originalsystem} is QSR-dissipative with storage function $V(x(k))= x(k)^\top P x(k)$ if and only if there exists a matrix $P=P^\top \geq 0,$ such that
\begin{equation}\label{eq:LMI_cond_model_based}
    \begin{bmatrix}
     A^\top PA-P-  C^\top QC & A^\top PB-C^\top S\\
     (A^\top PB-C^\top S)^\top & B^\top PB-R
    \end{bmatrix}\leq 0.
\end{equation}
\end{theorem}

Now we introduce the concept of passivity indices, which are associated to a specific form of QSR-dissipativity and, depending on sign conditions, quantify the deviations from passivity and variants of it, see \cite[Ch.~6]{khalil2002nonlinear} and \cite[Ch.~2]{lozano2013dissipative}. 
\begin{definition}[Passivity indices]
The system~\eqref{eq:originalsystem} with $m = p$ has \emph{passivity indices} $(\rho,\nu) \in \mathbb{R}^2$ when the supply rate in~\eqref{eq:supply-qsr} is chosen as
\begin{equation}
    Q = -\rho I_m, \quad 
    S = \tfrac{1}{2} I_m, \quad 
    R = -\nu I_m.
    \label{eq:QSR-passivity-structure}
\end{equation}

If $\rho>0$ ($\nu>0$), the system exhibits an \emph{excess of output (input) passivity}; conversely, if $\rho<0$ ($\nu<0$), it shows a \emph{shortage of output (input) passivity}.
\end{definition}

In MIMO systems, different input–output channels may contribute unevenly to the energy exchange with the environment. Then, having scalar global indices $\rho$ and $\nu$ may bring undesired conservativeness, particularly when analyzing the interconnection with other systems. The establishment of one of our main contributions relies on the slightly more general notion of input-output \emph{channel-wise} passivity indices $(\rho_1,\ldots,\rho_m,\nu_1,\ldots,\nu_m)$,
 which originates from the system \eqref{eq:originalsystem} being QSR-dissipative with
\begin{equation}
\scalebox{0.9}{$
    Q = -\begin{bmatrix}
        \rho_1  & \cdots & 0 \\
        \vdots & \ddots & \vdots \\
        0         & \cdots & \rho_m
    \end{bmatrix}, \quad
    S = \tfrac{1}{2}I_m, \quad
    R = -\begin{bmatrix}
        \nu_1  & \cdots & 0 \\
        \vdots & \ddots & \vdots \\
        0       & \cdots & \nu_m
    \end{bmatrix},
$}
\label{eq:channelwise_QSR}
\end{equation}
for some real numbers $\rho_i\in \mathbb{R},$ $\nu_i\in \mathbb{R}$, $i=1,2,\ldots,m$.

The solvability of \eqref{eq:LMI_cond_model_based} hinges upon the precise knowledge of the coefficient matrices $A$, $B$ and $C$. 
In the following subsection, inspired by \cite{alsalti2023notes,koch2022robust}, we present a data-based condition that only requires measured input–output data.

\subsection{Data-Based Dissipativity Characterization of MIMO LTI Systems}
Consider once again \eqref{eq:originalsystem} and henceforth assume that $A$, $B$ and $C$ are unknown. Let $u_{[0,T]}$ and $y_{[0,T]}$ denote measured input and output trajectories over a discrete-time horizon $T \in \mathbb{N}$, respectively. To assess whether \eqref{eq:originalsystem} is QSR-dissipative,
 we follow \cite{alsalti2023notes} to characterize a minimal realization of \eqref{eq:originalsystem} directly from the measured input-output data. 

Let $\ell$ denote the lag of \eqref{eq:originalsystem}, which is defined as the smallest integer such that $\text{rank}(O_{\ell})=n$, where \begin{equation*}
    O_{\ell}=\begin{bmatrix}
    C^{\top} & \left(CA\right)^{\top} & \cdots & \left(CA^{\ell-1}\right)^{\top}
\end{bmatrix}^{\top} \in \mathbb{R}^{p\ell \times n}.
\end{equation*}
Assume that $\ell \leq n \leq p\ell$~\citep{alsalti2023notes} (see also \cite{from_time_willems_1986}). 
We also require the following standard condition on the input signal~\citep{willems2005note}.

\begin{definition}[Persistency of excitation]
The input sequence $u_{[0,T]}$ is \emph{persistently exciting} of order $\ell+n+1$ if the Hankel matrix constructed from $u_{[0,T]}$ has full row rank.
\end{definition}

Then, in accordance to \cite[Lemma~1]{alsalti2023notes}, if $u_{[0,T]}$ is persistently exciting, then
    \begin{equation}
        \operatorname{rank} \left( \begin{bmatrix}
            U_{\{-\ell,\ell+1,T-1\}} \\[2pt]
            Y_{\{-\ell,\ell,T-2\}}
        \end{bmatrix}\right)= m(\ell+1)+n.
        \label{eq:rankcondition}
    \end{equation}
   Here, $U_{\{-\ell,\ell+1,T-1\}}$ and $Y_{\{-\ell,\ell,T-2\}}$ denote the Hankel matrices constructed from the input and output trajectories $u_{[0,T]}$ and $y_{[0,T]}$, respectively, following the notation introduced in Section~\ref{sec:data_based}.

Now let $z(k) \in \mathbb{R}^{m\ell + n}$ be a vector defined in terms of the measured data as follows:
\begin{equation}
    z(k) = \begin{bmatrix}
        u_{[k-\ell,k-1]} \\
        \overline{\Theta} \, y_{[k-\ell,k-1]}
    \end{bmatrix},
    \label{eq:zt_mimo}
\end{equation}
where $\overline{\Theta} \in \mathbb{R}^{n \times p\ell}$ is obtained by selecting $n$ linearly independent rows of $Y_{\{-\ell,\ell,T-2\}}$. This can always be done due to \eqref{eq:rankcondition} and the fact that $u_{[0,T]}$ is persistently exciting of order $\ell+n+1$ and thus $\text{rank}( U_{\{-\ell,\ell+1,T-1\}})=m(\ell+1)$.
In \cite{alsalti2023notes} it has been shown that $z(k)$ is a non-minimal state vector for \eqref{eq:originalsystem}, i.e., there exists a full row rank matrix $\hat T\in \mathbb{R}^{n\times (m\ell+n)}$, such that $x(k)=\hat{T} z(k)$ for all $k\in \mathbb{Z}$.
Then, there exist (unknown) matrices $\mathcal{A}$, $\mathcal{B}$ and $\mathcal{C}$, such that
\begin{equation}
   \begin{aligned}
        z(k+1)&=\mathcal{A} z(k)+\mathcal{B}u(k),\\
    y(k)&= \mathcal{C}z(k).
   \end{aligned}
   \label{eq:system_extended}
\end{equation}
The system~\eqref{eq:system_extended} has the same input-output behavior as \eqref{eq:originalsystem}. We present next how, subject to \eqref{eq:rankcondition}, it is possible to produce directly from $u_{[0,T]}$ and $y_{[0,T]}$ an equivalent data-based representation of~\eqref{eq:system_extended}.

Consider the system \eqref{eq:system_extended} and the original input sequence $u_{[0,T]}$. By computing the trajectory sequence $z_{[0,T]}$ in \eqref{eq:zt_mimo} and recalling the notation introduced in \eqref{eq:Z}, it is then possible to define the following (input-state) Hankel matrix associated to \eqref{eq:system_extended}:
\begin{equation}\label{eq:hankel_input_state_nonmin}
    H(u,z)= \begin{bmatrix}
        U_{\{0,T-1\}}\\
        Z_{\{0,T-1\}}
    \end{bmatrix} \in \mathbb{R}^{(m(\ell+1)+n) \times T}.
\end{equation}
In \cite{alsalti2023notes}, by establishing a linear relationship between $H(u,z)$ and the (input-output) Hankel matrix on the left-hand side of \eqref{eq:rankcondition}, it was shown that  $\text{rank}(H(u,z))=m(\ell +1)+n$. Therefore, it is possible to state the following result.

\begin{lemma}\label{lem:data_repr_z}
    Assume that $u_{[0,T]}$ is persistently exciting of order $\ell +n+1$. Then, the non-minimal representation \eqref{eq:system_extended} has the following equivalent data-based representation:
    \begin{equation}\label{eq:data_based_rep_nonmin}
    \begin{aligned}
        z(k+1) & = Z_{\{1,T\}} \begin{bmatrix}
            U_{\{0,T-1\}} \\[3pt]
            Z_{\{0,T-1\}}
        \end{bmatrix}^{\dagger} \begin{bmatrix}
            u(k) \\ z(k)
        \end{bmatrix},\\
      y(k) & =  Y_{\{0,T-1\}} Z_{\{0,T-1\}}^\dagger z(k), 
        \end{aligned}
    \end{equation}
where, $U_{\{0,T-1\}}$ and $Y_{\{0,T-1\}}$ follow the Hankel notation used in \eqref{eq:rankcondition}, while $Z_{\{0,T-1\}}$ is the corresponding Hankel matrix constructed from the trajectory $z_{[0,T]}$ as in \eqref{eq:hankel_input_state_nonmin}.
\end{lemma}
\begin{pf}
The proof can be established following a procedure analogous to the one used in \cite[Theorem~1]{depersis2019formulas} and is omitted due to space constraints.
\end{pf}

Moving forward, we can obtain the coefficient matrices $\mathcal{A}$, $\mathcal{B}$ and $\mathcal{C}$ that define \eqref{eq:system_extended} as follows~(c.f., \cite{koch2022robust}):
\begin{equation}
   \begin{split}
        \mathcal{A} &= \operatorname{Proj}_{m+1:(m+p)n+m} \left(Z_{\{1,T\}} H(u,z)^\dagger \right),
    \\
    \mathcal{B} &= \operatorname{Proj}_{1:m} \left(Z_{\{1,T\}}  H(u,z)^\dagger \right),\\
    \mathcal{C} &= Y_{\{0,T-1\}} Z_{\{0,T-1\}}^\dagger , 
   \end{split}
   \label{eq:def_XUH}
\end{equation}
where $\operatorname{Proj}_{a:b}(\cdot)$ denotes the selection of rows $a$ to $b$ from the given matrix. Using Kalman decomposition \citep[Theorem~6.7]{chen1984linear}, it is possible to use \eqref{eq:def_XUH} to construct a minimal state-space realization which is \emph{equivalent} to \eqref{eq:originalsystem}. Let $\hat{A}_i$, $\hat{B}_i$, and $\hat{C}_i$ denote the corresponding minimal realization system matrices, defined as
\begin{equation}\label{eq:minreal_subsystems}
\hat{A}_i = T_{\mathrm{co},i}^\top \mathcal{A}_i T_{\mathrm{co},i}, \qquad
\hat{B}_i = T_{\mathrm{co},i}^\top \mathcal{B}_i, \qquad
\hat{C}_i = \mathcal{C}_i T_{\mathrm{co},i},
\end{equation}
where $(\mathcal{A}_i,\mathcal{B}_i,\mathcal{C}_i)$ are defined as in~\eqref{eq:def_XUH}, and 
$T_{\mathrm{co},i}$ is an orthogonal matrix whose columns span the controllable and observable subspace\footnote{The controllable and observable subspace of a state-space triple $(A,B,C)$ is defined as 
$\mathcal{R}\cap \mathcal{U}^\perp$, where $\mathcal{R}$ is the column space of the controllability matrix and $\mathcal{U}$ is the kernel of the observability matrix~\citep{chen1984linear}.}.
Then, we have the following result.
\begin{theorem}\label{theorem:QSR-dissipativity}
      Assume that $u_{[0,T]}$  is persistently exciting of order $\ell +n+1$ and consider the matrices introduced in \eqref{eq:def_XUH}. Then, \eqref{eq:originalsystem} is QSR-dissipative if and only if there exists a matrix $\hat P = \hat P^{\top} \geq 0$, such that
    \begin{equation}
    \begin{bmatrix}
        \hat{A}^{\top} \hat P \hat{A} - \hat P - \hat{C}^{\top}Q\hat{C} 
        & 
        \hat{A}^{\top} \hat P \hat{B} - \hat{C}^{\top}S 
        \\[4pt]
        (\hat{A}^{\top} \hat P \hat{B} - \hat{C}^{\top}S)^{\top}
        & 
        \hat{B}^{\top} \hat P \hat{B} - R
    \end{bmatrix}
    \leq 0,
    \label{eq:QSR-DissipativityData} 
\end{equation}
where $Q$, $S$ and $R$ are as in \eqref{eq:channelwise_QSR}.
\end{theorem}
\begin{pf}
The proof is a direct application of Theorem~\ref{theorem:qsr_dissip_model_based} after noting that \eqref{eq:QSR-DissipativityData} implies QSR-dissipativity of the minimal system realization $\hat{x}(k+1)=\hat A \hat x(k)+\hat B u$, $y=\hat C \hat x$, which is equivalent to \eqref{eq:originalsystem}.
\end{pf}
\section{Problem Statement and Main Result}

Having established the necessary background in the previous section, we are in position to formulate the problem statement and present our main result.

\subsection{Problem Statement}

Consider a network of $N>1$ interconnected subsystems $\Sigma_i$, whose dynamics are discrete-time LTI MIMO of the form:
\begin{equation}   \label{eq:linear_subsystem}
\Sigma_{i}: \quad 
\begin{aligned}
        x_{i}(k+1) & =  A_{i}\,x_{i}(k) + B_{i}\,u_{i}(k), \\[2pt]
        y_{i}(k) & = C_{i}\,x_{i}(k),
  \end{aligned}
\end{equation}
where $x_i(k) \in \mathbb{R}^{n_i}$, $u_i(k) \in \mathbb{R}^{m_i}$ and $y_i(k) \in \mathbb{R}^{m_i}$. We assume that the matrices $A_i$, $B_i$, and $C_i$ are unknown, and that each subsystem $\Sigma_i$ is controllable and observable. Moreover, let $M_i = \{1,2,\dots,m_i\}$ be the index set of input-output pairs for the $i$-th subsystem. For each $j \in M_i$, we assume that there exists a unique pair $(a,b)$, with
\begin{equation}
a \in \{1,2,\dots,N\} \setminus \{i\}, \quad b \in M_a,
\end{equation}
such that the interaction among the two subsystems $i$ and $a$ is carried out through  the following feedback interconnection law of their $j$-th and $b$-th channel:
\begin{equation}
\begin{aligned}
    u_{i,j} &= y_{a,b}, \quad
    u_{a,b} = - y_{i,j}.
\end{aligned}
\label{eq:interconnection}
\end{equation}
Then, the main problem we address in this paper is as follows: Assuming that input-output measurement sets, $u_{[0,T_i]}$ and $y_{[0,T_i]}$, are locally available to its corresponding subsystem $\Sigma_i$, derive data-based and tractable compositional conditions to assess whether the overall interconnected system \eqref{eq:linear_subsystem}, \eqref{eq:interconnection}, $i=1,\ldots,N$, is stable.


\subsection{Main Result}

As our main result, we present compositional data-based conditions for certifying the stability of interconnected discrete time LTI MIMO systems, subject to the availability of sufficiently rich local input-output data. Conditions for asymptotic stability are given in Corollary~\ref{cor:AS}.

\begin{theorem}\label{theorem:main_result}
Consider the interconnected system \eqref{eq:linear_subsystem}, \eqref{eq:interconnection}, $i=1,\ldots,N$. For each subsystem $\Sigma_i$ assume that input-output data sequences $u_{[0,T_i]}$ and $y_{[0,T_i]}$ are available to each subsystem, where $u_{[0,T_i]}$ is persistently exciting of order $\ell_{i} +n_{i}+1$. Then, the origin of the interconnected system is stable if there exist matrices $\hat P_i=\hat P_i^\top>0$, such that 
    \begin{subequations}
\begin{align}
    \begin{bmatrix}
        \hat{A}_i^{\top} \hat P_i \hat{A}_i - \hat P_i - \hat{C}_i^{\top}Q_i\hat{C}_i 
        & 
        \hat{A}_i^{\top} \hat P_i \hat{B}_i - \hat{C}_i^{\top}S_i 
        \\[4pt]
        (\hat{A}_i^{\top} \hat P_i \hat{B}_i - \hat{C}_i^{\top}S_i)^{\top}
        & 
        \hat{B}_i^{\top} \hat P_i \hat{B}_i - R_i
    \end{bmatrix}
    & \leq 0, \label{eq:LMI_subsystems_QSR}\\
\begin{bmatrix}
    -\rho_{i,j} -\nu_{a,b} & 0 \\
    0 & -\rho_{a,b} -\nu_{i,j}
\end{bmatrix} & \leq 0, \label{eq:LMI_subsys_STAB}
\end{align}
\end{subequations}
where
\begin{equation}\label{eq:QSR_matrices_subsystems}
\begin{aligned}
    Q_i &= \operatorname{diag}(-\rho_{i,1}, -\rho_{i,2}, \dots, -\rho_{i,m_i}), \\
    R_i &= \operatorname{diag}(-\nu_{i,1}, -\nu_{i,2}, \dots, -\nu_{i,m_i}), \\
    S_i &= \frac{1}{2} I_{m_i \times m_i},
\end{aligned}
\end{equation}
where $(\hat{A}_i,\hat{B}_i,\hat{C}_i)$ is given in \eqref{eq:minreal_subsystems}.
\end{theorem}
\begin{pf}
By  Theorem~\ref{theorem:QSR-dissipativity} the existence of $\hat{P}_i>0$ satisfying \eqref{eq:LMI_subsystems_QSR} implies that each subsystem $\Sigma_i$ is QSR-dissipative with positive-definite quadratic storage function $V_i(x_i(k))=\frac{1}{2}x_i^\top(k)P x_i(k)$ and with respect to the supply rate
\begin{equation}
    w_i(u_i(k),y_i(k))= \begin{bmatrix}
        y_i(k) \\ u_i(k)
    \end{bmatrix}^\top
    \begin{bmatrix}
        Q_i & S_i \\
        S_i^\top & R_i
    \end{bmatrix}
    \begin{bmatrix}
        y_i(k) \\ u_i(k)
    \end{bmatrix}.
\end{equation}
Let $V(x(k))=\sum_{i=1}^N V_i(x_i(k))$ be a candidate Lyapunov function for showing that the origin of the interconnected system \eqref{eq:linear_subsystem}, \eqref{eq:interconnection}, $i=1,\ldots,N$, is stable. Then, due to QSR-dissipativity of each $\Sigma_i$ it holds that
\begin{equation}\label{eq:time_deriv_aggregated}
      V(x(k+1))-V(x(k))\leq \sum_{i=1}^Nw_i(u_i(k),y_i(k)).
\end{equation}
We write now an equivalent expression for the right-hand side of \eqref{eq:time_deriv_aggregated}. Considering the  structures for $Q_i$, $S_i$ and $R_i$ in \eqref{eq:QSR_matrices_subsystems}, we obtain from \eqref{eq:supply-qsr} that
\begin{equation}\label{eq:dissp_inequality}
     \sum_{i=1}^{N} w_i(u_i(k),y_i(k)) 
= \sum_{i=1}^{N} \sum_{j \in M_i} \left(-\rho_{i,j} y_{i,j}^2 + y_{i,j} u_{i,j} -\nu_{i,j} u_{i,j}^2 \right),
\end{equation}
where we recall that $M_i = \{1,2,\dots,m_i\}$ is the index set of input-output pairs in subsystem $\Sigma_i$. Now, due to the interconnection law \eqref{eq:interconnection}, it holds that each term $(y_{i,j}, u_{i,j})$ has a \emph{unique} counterpart $(y_{a,b}, u_{a,b})$, such that the corresponding two elements in \eqref{eq:dissp_inequality} are given by 
\begin{equation*}
\begin{aligned}
&-\rho_{i,j} y_{i,j}^2 + y_{i,j} y_{a,b} - \nu_{a,b} y_{i,j}^2 
-\rho_{a,b} y_{a,b}^2 - y_{a,b} y_{i,j} -\nu_{i,j} y_{a,b}^2 \\
&= -(\rho_{i,j} +\nu_{a,b}) y_{i,j}^2 - (\rho_{a,b} +\nu_{i,j}) y_{a,b}^2.
\end{aligned}
\end{equation*}
Let $\Pi_i$ be the set of all pairs $(a,b)$ such that the channels $(i,j)$ and $(a,b)$ are interconnected according to \eqref{eq:interconnection}. 
For each subsystem $k \neq i$, we define
\begin{equation}
    \mathcal{K}_{k,i} := \{\, j \in M_i \;\mid\; \exists\, b \in M_k \text{ such that } (k,b) \in \Pi_i \,\}.
    \label{eq:def_Kki}
\end{equation}
This set collects the input-output channels of $\Sigma_i$ that are interconnected with subsystem $\Sigma_k$.
Therefore, the following identities can be obtained:
\begin{align}
&\sum_{i=1}^Nw_i(u_i(k),y_i(k))= \nonumber \\
&\sum_{i=1}^{N} \sum_{j \in M_i} \left(-\rho_{i,j} y_{i,j}^2 + y_{i,j} u_{i,j} -\nu_{i,j} u_{i,j}^2 \right)= \nonumber\\
& \sum_{i=1}^{N-1} \sum_{j \in M_i \setminus \bigcup_{k=1}^{i-1} \mathcal{K}_{k,i}}
\begin{bmatrix}
y_{i,j} \\
y_{a,b}
\end{bmatrix}^\top
\begin{bmatrix}
-\rho_{i,j} - \nu_{a,b} & 0 \\
0 & -\rho_{a,b} -\nu_{i,j}
\end{bmatrix}
\begin{bmatrix}
y_{i,j} \\
y_{a,b}
\end{bmatrix}. \label{eq:sum_ws}
\end{align}
The union $\bigcup_{k=1}^{i-1} \mathcal{K}_{k,i}$ ensures that each interconnection pair is counted only once.
In view of \eqref{eq:LMI_subsys_STAB} and \eqref{eq:time_deriv_aggregated}, we obtain that
\begin{equation}
          V(x(k+1))-V(x(k))\leq 0,
\end{equation}
which implies that $V$ is a Lyapunov function for the origin of the interconnected system and, consequently, the origin is stable.
\end{pf}

The next two corollaries follow from Theorem~\ref{theorem:main_result}.

\begin{corollary}\label{cor:AS}
Consider the interconnected system \eqref{eq:linear_subsystem}, \eqref{eq:interconnection}, $i=1,\ldots,N$.
     Assume that \eqref{eq:LMI_subsys_STAB} holds with strict inequality. Then, the origin of the overall interconnected system is asymptotically stable provided that every subsystem $\Sigma_i$ is either zero state observable or zero state detectable. 
\end{corollary}
\begin{pf}
From Theorem~\ref{theorem:main_result}, any $\alpha$-sublevel set $\Omega_\alpha$ 
of the Lyapunov function $V=\sum_{i=1}^N\frac{1}{2}x_i^\top(k)P x_i(k)$ 
is invariant (with $\alpha>0$ arbitrary). Moreover, \eqref{eq:time_deriv_aggregated} and \eqref{eq:sum_ws} hold.

Since \eqref{eq:LMI_subsys_STAB} holds strictly, it follows that 
$ V(x(k+1))-V(x(k))=0$  
if and only if $y_{i,j}=0$ for every 
$i=1,\dots,N$ and every $j \in M_i$. 
To make this explicit, define the set 
$\mathcal{E} := \{(x_1,\dots,x_N): y_{i,j}=0 \ \forall\, i,\, j\in M_i\}$, 
which characterizes the states for which $V$ remains constant.
If each $\Sigma_i$ is zero-state observable, i.e., 
$y_{i,j}=0$ for all $i,j$ implies $x_i= 0$, then $V$ becomes a strict Lyapunov function.
In this case, $\mathcal{E}=\{0\}$. Alternatively, if each $\Sigma_i$ is zero-state detectable, i.e., 
$y_{i,j}=0$ for all $i,j$ implies $x_i\rightarrow 0$, then 
the origin becomes the largest attractive invariant set contained in 
$\{(x_1,\dots,x_N)\in \Omega_\alpha:~ y_{i,j}=0~\forall j\in M_i\}$. 
Equivalently, the origin is the largest attractive invariant set contained in $\mathcal{E}\cap \Omega_\alpha$.
By LaSalle's invariance principle 
\citep{mei_bullo2017lasalle}, 
for discrete-time systems, 
the origin is asymptotically stable.
\end{pf}
\begin{corollary}\label{cor:opt}
Consider the interconnected system \eqref{eq:linear_subsystem}, \eqref{eq:interconnection}.  
For each subsystem $\Sigma_i$, its channel-wise passivity indices $(\rho_{i,j},\nu_{i,j})$ can be obtained as the solution of the optimization problem
\begin{equation}
    \max_{\rho_{i,j}, \nu_{i,j}} \; f(\rho_{i,j},\nu_{i,j}),
    \label{eq:convex_opt}
\end{equation}
subject to the local QSR-dissipativity condition~\eqref{eq:LMI_subsystems_QSR} and the stability condition~\eqref{eq:LMI_subsys_STAB} for all relevant pairs involving $(\rho_{i,j},\nu_{i,j})$.  
Moreover, the computation can be carried out in a distributed manner, in the sense that each subsystem enforces~\eqref{eq:LMI_subsystems_QSR} using its own input--output data, while the constraints~\eqref{eq:LMI_subsys_STAB} only couple indices across directly interconnected subsystems.
\end{corollary}

\begin{pf}

Both constraints \eqref{eq:LMI_subsystems_QSR} and \eqref{eq:LMI_subsys_STAB} are linear matrix inequalities in $(\hat{P}_i,\rho_{i,j},\nu_{i,j})$, and are therefore convex constraints. Hence, if the cost function $f$ in \eqref{eq:convex_opt} is convex, the optimization problem is convex.  
Any feasible solution satisfies the imposed dissipativity and stability constraints involving $(\rho_{i,j},\nu_{i,j})$ for each subsystem and its neighbors.
\end{pf}

\section{Numerical Example}

In this section, for a low-voltage DC microgrid we illustrate the proposed data-based dissipativity analysis. Using measured input--output trajectories of four distinct system areas, we evaluate local QSR-dissipativity properties before and after  generation-outage contingencies, and illustrate the associated changes in the channel-wise passivity indices and the implication for the verification of the compositional stability condition introduced in Section~III.

\subsection{Simulation Setup}

\begin{figure*}[!t]
    \centering
    \includegraphics[width=\linewidth,trim=30 5 30 5,clip]{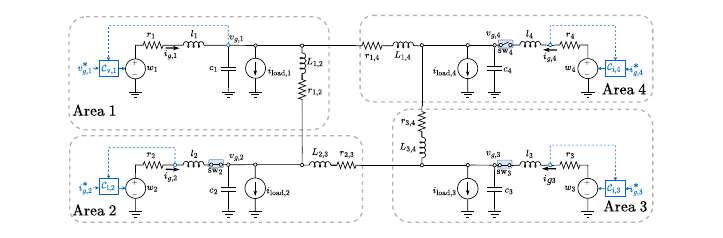}
   \caption{Low-voltage DC microgrid used in the numerical study. The four areas (gray dashed boxes) are interconnected by the power lines $\{1,2\}$, $\{2,3\}$, $\{3,4\}$, and $\{1,4\}$ with parameters $(r_{ij},L_{ij})$. In each area, $(r_i,l_i)$ models the converter filter, $c_i$ the local bus capacitor, $v_{g,i}$ the bus voltage, and $i_{\text{load},i}$ the aggregated load. All loads are of ZI type. Area~1 operates in voltage-setting mode (PI controller $\mathcal C_{\mathrm{v},1}$ regulates $v_{g,1}$ towards a reference), whereas Areas~2–4 operate in current-setting mode (PI controllers regulate $i_{g,i}$ towards a reference). The local switches $\mathtt{sw}_i=2,3,4$, can be opened to emulate $N\!-\!1$ generation-outage contingencies.}
    \label{fig:microgrid_setup}
\end{figure*}

The considered microgrid setup shown in Fig.~\ref{fig:microgrid_setup} is a variant of the case study reported in \cite{cucuzzella_robust_tcst_2019}, which consists of four main areas, each with local power generation and consumption, and which are interconnected through resistive-inductive power lines. For reproducibility, the main parameters and configuration used to generate the input–output data are reported in Appendix~\ref{appendix:modeling}, although the data-based analysis does not rely on explicit model knowledge.
We consider the following three generation–outage contingencies: at $t=0$~s the microgrid is initialized at a stable steady-state and remains so until $t=25$~s, when a single contingency is simulated through the opening $\mathtt{sw}_i$, which implies the disconnection of the generation unit of Area $i=2,3,4$, see~Fig.~\ref{fig:microgrid_setup}. 

The data-driven workflow is illustrated using the outage of generation unit $i=4$ as the baseline scenario. For each scenario, the generation unit is assumed to remain disconnected for the remainder of the simulation\footnote{For the discrete-time analysis, we use the zero-order hold (\texttt{zoh}) sampled-data realization with period $T_s=3.4\times10^{-4}\,$s.}. Moreover, the interconnections among the areas obey the following relations, which are consistent with the interconnection law \eqref{eq:interconnection}:
\begin{equation}
    \begin{aligned}
        u_1 & = \begin{bmatrix}
           -y_{4,2} & y_{2,1}\\
        \end{bmatrix}^\top, &\quad  u_3 & = \begin{bmatrix}
            -y_{2,2} &y_{4,1}
        \end{bmatrix}^\top,\\
        u_2 & = \begin{bmatrix}
             -y_{1,2} & y_{3,1}
        \end{bmatrix}^\top, &\quad
        u_4 & = \begin{bmatrix}
           -y_{3,2}& y_{1,1}
        \end{bmatrix}^\top.
    \end{aligned}
    \label{eq:relations_simu}
\end{equation}
Considering the definition of the coefficient matrices $C_{i}$ in \eqref{eq:pre_fault_area_1} and \eqref{eq:Area2-4_prefaultsystem}, the measured output of each area is given by $y_i=\begin{bmatrix}v_{g,i} & i_{i,j}\end{bmatrix}^\top$. For each $i$, the input $u_i=[u_{i,1}\;u_{i,2}]^\top$ is comprised of the current $u_{i,1}$ of one of the power lines incident to Area $i$, and the bus voltage $u_{i,2}$ at the other end bus of the line $\{i,j\}$ already embedded in Area $i$. Thus, the relations in \eqref{eq:relations_simu} satisfy those in \eqref{eq:interconnection}. 

\subsection{Numerical Results}
In order to carry out the proposed data-driven analysis, we assume that basic information about the composition of each microgrid area is available: each area is known to have a converter, a local ZI load, an RL line, and a PI controller. Hence, each area has four dynamic components. Consequently, we infer for the input-output data-driven analysis that each area has to be represented by a fourth-order system ($n=4$) with two input-output pairs. Since we have two measured outputs per area, we set the Hankel lag to $\ell=2$, satisfying $\ell \leq 4 \leq 2\ell$ (see Section~\ref{sec:data_based}). After a fault, the disconnected area behaves as a second-order system ($n=2$, $\ell=1$), while other areas remain fourth-order ($n=4$, $\ell=2$).

\begin{figure}[ht]
    \centering
    \includegraphics[width=0.9\linewidth]{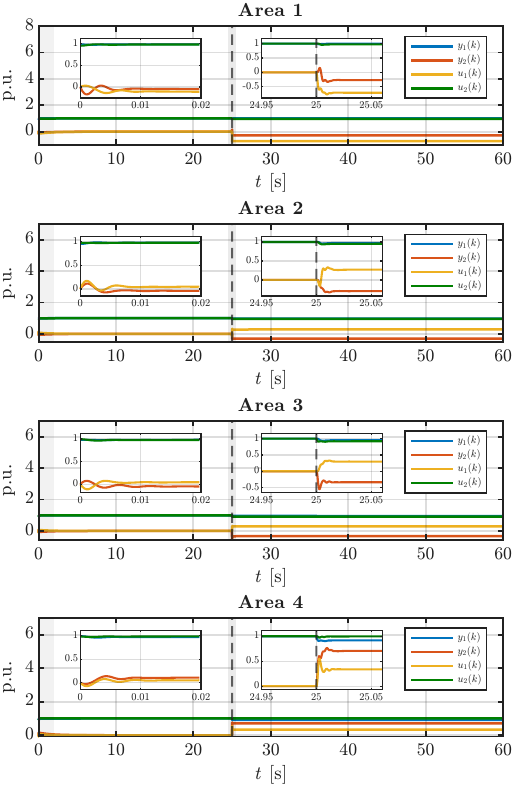}
    \caption{Pre-fault and Post-fault trajectories of $u_i(t)$ and $y_i(t)$ (p.u.) in Areas~1–4. Insets show zoom over $[0,0.02]$~s and $[24.95,25.05]$~s}.
    \label{fig:evolution_areas}
\end{figure}

Figure~\ref{fig:evolution_areas} shows the system evolution from $t=0$~s to $t=60$~s for the baseline scenario in which the generation unit of Area~4 is disconnected at $t=25$~s.

The proposed data-based analysis is independently implemented both on the pre-fault and the post-fault conditions. For both conditions we first verify that the measured input-output data meet the rank condition \eqref{eq:rankcondition}, ensuring persistent excitation of the measured data for the model order $n$ and Hankel lag $\ell$. Then, for each area we construct a non-minimal data-based representation (Lemma~\ref{lem:data_repr_z}), from which we obtain a minimal realization as in \eqref{eq:minreal_subsystems}. The minimal realization models are then used to evaluate the QSR-dissipativity condition in Theorem~\ref{theorem:QSR-dissipativity} and compute the channel-wise passivity indices needed for verifying the stability condition in Theorem~\ref{theorem:main_result}. We note that the interconnection law \eqref{eq:interconnection} holds in this example, and we can thus solve a convex optimization problem for computing the channel-wise passivity as outlined in Corollary~\ref{cor:opt}: the optimization problem is solved using YALMIP, see ~\cite{yalmip_lofberg}, and considering the cost function $f(\rho)=\min(-\sum_j \rho_{i,j})$ and a $10^{-3}$ numerical tolerance on the stability constraints.

The resulting channel–wise passivity indices are summarized in Table~\ref{tab:passivity_comparison}. Across all areas, every channel exhibits an excess of output passivity (all $\rho_{i,j} > 0$) and a shortage of input passivity (all $\nu_{i,j} < 0$). In the pre-fault window, Areas~1 and~4 exhibit the largest passivity margins, whereas Areas~2 and~3 show the smallest ones, with $\rho_{2,2}$ and $\rho_{3,1}$ being the lowest entries. After the fault, several channels in Areas~1, 2, and~4 experience reductions, with the most notable decreases in $\rho_{1,1}$ and $\rho_{2,2}$. Although all $\rho_{i,j}$ indices remain positive, the post-fault values indicate a reduced excess of output passivity in the affected areas, while the compositional stability condition of Theorem~\ref{theorem:main_result} continues to hold.

\begin{table}[ht]
  \centering
  \scriptsize
  \caption{Comparison of pre- and post-fault channel-wise passivity indices.}
  \label{tab:passivity_comparison}
  \setlength{\tabcolsep}{3.5pt}
  \begin{tabular}{@{}ccccccccc@{}}
    \toprule
    Area 
    & $\rho_{i,1}^{\text{pre}}$ & $\rho_{i,1}^{\text{post}}$
    & $\rho_{i,2}^{\text{pre}}$ & $\rho_{i,2}^{\text{post}}$
    & $\nu_{i,1}^{\text{pre}}$  & $\nu_{i,1}^{\text{post}}$
    & $\nu_{i,2}^{\text{pre}}$  & $\nu_{i,2}^{\text{post}}$ \\
    \midrule
    1 & 0.6004 & 0.5654 & 0.6098 & 0.6366 & -1.2000 & -1.0460 & -0.4421 & -0.4388 \\
    2 & 0.4431 & 0.4398 & 0.5666 & 0.5155 & -0.6088 & -0.6356 & -0.4079 & -0.4095 \\
    3 & 0.4089 & 0.4105 & 0.8618 & 0.8550 & -0.5056 & -0.5145 & -0.5177 & -0.4940 \\
    4 & 0.5187 & 0.4950 & 1.2012 & 1.0470 & -0.8608 & -0.8540 & -0.5994 & -0.5644 \\
    \bottomrule
  \end{tabular}
\end{table}

We also evaluated pre–fault and post–fault dissipativity for contingencies in Areas~2 and~3, where the corresponding generation unit is disconnected at $t=25$\,s. In these cases, distinct PI gains were selected for each scenario and kept fixed throughout the simulation, since other gain choices led to data that either violated the persistency of excitation condition or produced unreliable data-based realizations due to discretization and sampling effects. The parameter sets in Table~\ref{tab:n1_summary} yield informative trajectories and consistent identified models in both time windows. With these parametrizations, the channel-wise passivity indices display the same qualitative trends as in the baseline case, so that we don't report them here explicitly. Likewise, the post–fault equilibria in all scenarios satisfy the compositional stability condition of Theorem~\ref{theorem:main_result}.

\begin{table}[ht]
  \centering
  \caption{PI control gains for the fault scenarios in Areas~2~and~3.}
  \label{tab:n1_summary}
  \begin{tabular}{c c c | c c c c}
   Scenario&~ & ~ &Area~1 & Area~2 & Area~3 & Area~4 \\
    \hline
    &$k_{\text{prop}}$   & (--)      & 1.5 & 1 & 11 & 18 \\
   $\mathtt{sw}_2$ open &$k_{\text{integ}}$  & (s$^{-1}$)& 100 & 1 &  9 & 12 \\
    \hline
    &$k_{\text{prop}}$   & (--)      & 1.5 & 29          & 1 & 18 \\
    $\mathtt{sw}_3$ open &$k_{\text{integ}}$  & (s$^{-1}$)& 100 & 28          & 1 & 16 \\
  \end{tabular}
\end{table}

\section{Conclusion and Future Work}

We have presented a compositional data-driven method for assessing the stability of interconnected MIMO LTI systems based solely on input-output measurement data of each of the component subsystems. Our  approach relies on the data-based characterization of non-minimal state-space  representations of all the component subsystems. This allows us to formulate structured LMIs for verifying  QSR-dissipativity properties of the subsystems and to assess the stability of the interconnected system in a compositional manner. The effectiveness of our approach was tested on a numerical example of a DC microgrid facing faults.

In future work, we will investigate local stability properties of networks of interconnected nonlinear systems. Moreover, we will explore control gain tuning approaches that can support the verification of persistency of excitation conditions on the measured data. In addition, we will expand our methodology to deal with experimental/field and noise-corrupted measurements.

\begin{ack}
This research is supported by the German Federal Government, the Federal Ministry of Research, Technology and Space and the State of Brandenburg within the framework of the joint project EIZ: Energy Innovation Center (project numbers 85056897 and 03SF0693A) with funds from the Structural Development Act (Strukturstärkungsgesetz) for coal-mining regions.
\end{ack}

\bibliography{ifacconf}           

\appendix
\section{Modeling setup}\label{appendix:modeling}
\begin{figure}[!t]
    \centering
     \begin{subfigure}{0.48\linewidth}
        \centering
        \includegraphics[width=\linewidth,trim=30 5 28 10,clip]{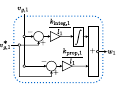}
        \caption{}
        \label{fig:controllers:forming}
    \end{subfigure}
    \hfill
    \begin{subfigure}{0.48\linewidth}
        \centering
        \includegraphics[width=\linewidth,trim=28 5 28 10,clip]{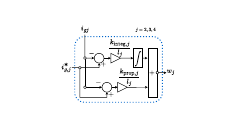}
        \caption{}
        \label{fig:controllers:following}
    \end{subfigure}
    \caption{Controller setup (a) Voltage-setting PI controller  for Area 1, (b) Current-setting PI controller for Areas 2,3 and 4.}
    \label{fig:controllers}
\end{figure}

The (continuous-time) pre-contingency dynamics of each area is described by a model of the form of \eqref{eq:linear_subsystem} plus a constant additive term $d_i$, and with the following coefficient matrices:
\begin{equation}\label{eq:pre_fault_area_1}
\begin{aligned}
    A_1 & =\begin{bmatrix}
        -\tfrac{r_1}{l_1} & -\tfrac{1}{l_1}(1+k_{\text{prop},1}) & \tfrac{1}{l_1} & 0\\
        \tfrac{1}{c_1} & -\tfrac{1}{c_1r_{\text{load},1}} & 0 & \tfrac{1}{c_{1}}\\
        0 & -k_{\text{integ},1} & 0 &   0\\
        0&-\tfrac{1}{L_{1,2}} &0 &-\tfrac{r_{1,2}}{L_{1,2}}
    \end{bmatrix},\\
    B_1 & =\begin{bmatrix}
        0 & 0\\
        \tfrac{1}{c_1} &   0\\
        0 & 0\\
         0 &   \tfrac{1}{L_{1,2}}
    \end{bmatrix},\quad  C_1  =\begin{bmatrix}
        0 & 1 & 0 & 0\\
        0 & 0 & 0 & 1
    \end{bmatrix},\\
    d_1 & =\begin{bmatrix}
        \tfrac{k_{\text{prop},1}x_{1,2}^\star}{l_1} & -\tfrac{i_{\text{load},1}}{c_1} & k_{\text{integ},1}x_{1,2}^\star & 0
    \end{bmatrix}^\top ,
\end{aligned}
\end{equation}
and, for $i=2,3,4$,
\begin{equation}
\label{eq:Area2-4_prefaultsystem}
\begin{aligned}
    A_i & =\begin{bmatrix}
        -\tfrac{1}{l_i}(r_i+k_{\text{prop},i}) & -\tfrac{1}{l_i} & \tfrac{1}{l_i}&0\\
        \tfrac{1}{c_i} & -\tfrac{1}{c_ir_{\text{load},i}} & 0&\tfrac{1}{c_{i}}\\
        -k_{\text{integ},i} & 0 & 0& 0\\
      0&-\tfrac{1}{L_{i,j}}&0 &-\tfrac{r_{i,j}}{L_{i,j}}
    \end{bmatrix},\\
    B_i & =\begin{bmatrix}
        0 & 0\\
        \tfrac{1}{c_i} & 0\\
        0 & 0 \\
      0 & \tfrac{1}{L_{i,j}}
    \end{bmatrix},\quad 
  C_i  =\begin{bmatrix}
        0 & 1 & 0 &0\\
        0 &0 & 0 & 1
    \end{bmatrix},\\
    d_i & =\begin{bmatrix}
        \tfrac{k_{\text{prop},i}x_{i,1}^\star}{l_i} & -\tfrac{i_{\text{load},i}}{c_i}& k_{\text{integ},i}x_{i,1}^\star &0
    \end{bmatrix}^\top.
\end{aligned}
\end{equation}
The considered numerical parameters are shown in Tables~\ref{tab:dgus_param} and \ref{tab:lines_param}. 
For each area, the components of the state vector $x_i=(x_{i,1},x_{i,2},x_{i,3},x_{i,4})$ represent current generation $i_{g,i}$, bus voltage $v_{g,i}$, an integral control action $z_{g,i}$, and the current $i_{i,j}$ of one of the power lines incident to Area $i$: in Areas 1 to 4 we respectively embed the lines $\{1,2\}$, $\{2,3\}$, $\{3,4\}$, and $\{1,4\}$, with parameters $(r_{i,j},L_{i,j})$; we set the convention that the reference direction for positive current $i_{i,j}$ is from $j$ to $i$. 
The goal of the generation unit at Area 1 is to establish, through a PI-control action (see Fig.~\ref{fig:controllers}), a desired steady-state bus voltage $x_{1,2}^\star$, whereas the goal of the generation units at Areas 2, 3 and 4 is to establish (also through PI-control) desired steady-state current injections $x_{2,1}^\star$, $x_{3,1}^\star$ and $x_{4,1}^\star$, respectively.
 \begin{table}
    \centering
    \caption{Microgrid Parameters and Desired Setpoints}
    \label{tab:dgus_param}
    \begin{tabular}{c c |c c c c}
        Area & ~ & 1 & 2 & 3 & 4   \\
        \hline
$r_i$ & ($\Omega$) & 0.2 & 0.3 & 0.5 & 0.1\\
$l_i$ & (mH) & 1.8 & 2.0 & 3.0 & 2.2\\
$c_i$ & (mF) & 2.2 & 1.9 & 2.5 & 1.7\\
$r_{\text{load},i}$ & ($\Omega$) & 7.70 & 12.84 & 12.84 & 9.63\\
$i_{\text{load},i}$ & (A) & 16.45 & 9.87 & 9.87 & 13.16\\
$x_{1,2}^\star$ & (V) & 380 & --   & --  & --\\
$x_{i,1}^\star$ & (A) & --   & 39.47 & 39.47 & 52.63\\
$k_{\text{prop}}$ &  (--)  & 2.1 & 19& 35&1\\
$k_{\text{integ}}$ & (s$^{-1}$)& 60& 11& 14&1\\
    \end{tabular}
\end{table}
\begin{table}[!t]
    \centering
    \caption{Power Line Parameters}
    \label{tab:lines_param}
    \begin{tabular}{c c |c c c c}
        Line & ~ & \{1,2\} & \{2,3\} & \{3,4\} & \{1,4\}   \\
        \hline
$r_{i,j}$ & ($\Omega$) & 0.70 & 0.60 & 0.80 & 0.90\\
$L_{i,j}$ & (mH)      & 0.8  & 1  & 1  & 0.7
    \end{tabular}
\end{table}
\end{document}